\documentclass[prd,aps,floats,floatfix,eqsecnum,nofootinbib]{revtex4}
\usepackage{verbatim,graphicx,amssymb,amsbsy,bm,amsmath,rotating,epsfig}
\usepackage{psfrag}
\usepackage{hyperref}
\usepackage{fontenc}
\newcommand{\be}{\begin{equation}}
\newcommand{\ee}{\end{equation}}
\newcommand{\bea}{\begin{eqnarray}}
\newcommand{\eea}{\end{eqnarray}}

\newcommand{\bp}{\ensuremath{\mathbf{p}}}

\newcommand{\br}{\ensuremath{\mathbf{r}}}

\begin{document}
\title{Dark matter in galaxies: the dark matter particle mass is about 7 keV \footnote{Based
on Lectures given by H J de V at NuMass 2013, Milano-Bicocca, February 2013; at Cosmic Frontiers,
SLAC, March 2013 and  by H J de V and N G S at the Chalonge Torino Colloquium 2013, April 2013 and The Paris Chalonge Colloquium July 2014}}
\author{\bf H. J. de Vega $^{(a)}$}
\email{devega@lpthe.jussieu.fr} 
\author{\bf N. G. Sanchez $^{(b)}$}
\email{Norma.Sanchez@obspm.fr} 
\affiliation{
$^{(a)}$ LPTHE, Universit\'e
Pierre et Marie Curie (Paris VI),
Laboratoire Associ\'e au CNRS UMR 7589, Tour 13, 4\`eme. et 5\`eme. \'etages, 
Boite 126, 4, Place Jussieu, 75252 Paris, Cedex 05, France. \\
$^{(b)}$ Observatoire de Paris, LERMA. Laboratoire Associ\'e au CNRS UMR 8112.
\\61, Avenue de l'Observatoire, 75014 Paris, France.}
\date{\today}
\begin{abstract}
Warm dark matter (WDM) means DM particles with mass $ m $ in the keV scale.
For large scales, (structures beyond $ \sim 100$ kpc) WDM and CDM yield identical results 
which agree with observations. For intermediate scales, WDM gives the correct abundance of substructures.
Inside galaxy cores, below $ \sim 100$ pc, $N$-body WDM classical physics simulations 
are incorrect because at such scales quantum WDM effects are important.
WDM quantum calculations (Thomas-Fermi approach) provide galaxy cores, 
galaxy masses, velocity dispersions and density profiles in agreement with the observations.
For a dark matter particle decoupling at thermal equilibrium (thermal relic),
all evidences point out to a 2 keV particle. Remarkably enough, sterile neutrinos decouple
out of thermal equilibrium with a primordial power spectrum similar
to a 2 keV thermal relic when the sterile neutrino mass is about 7 keV.
Therefore, WDM can be formed by 7 keV sterile neutrinos.
Excitingly enough, Bulbul et al. (2014) announced the detection of a cluster X-ray emission line 
that could correspond to the decay of a  7.1 keV sterile neutrino and to a neutrino decay mixing angle 
of $ \sin^2 2\; \theta \sim 7\; 10^{-11} $. This is a further argument in favour of sterile neutrino WDM.
Baryons, represent 10\% of DM or less in galaxies and are expected to give a correction to pure WDM results.
The detection of the DM particle depends upon the particle physics model.
Sterile neutrinos with keV scale mass (the main WDM candidate) can be detected in 
beta decay for Tritium and Renium and in the electron capture in Holmiun.
The sterile neutrino decay into X rays can be detected observing DM
dominated galaxies and through the distortion of the black-body CMB spectrum.

So far, {\bf not a single valid} objection arose against WDM.
\end{abstract}
\maketitle
\tableofcontents

\section{Introduction}

81 \% of the matter of the universe is {\bf dark}.
Dark matter (DM) is the dominant component of galaxies. 
DM interacts through gravity.

\medskip

DM interactions other than gravitational have been {\bf so far unobserved}, such possible couplings must be very weak:
much weaker than weak interactions in particle physics. DM is outside the standard model of particle physics.

\medskip

The main proposed candidates for DM are:
Neutrinos (hot dark matter) back in the 1980's with particle mass $ m \sim 1 $ eV
already ruled out, 
Cold Dark Matter (CDM), weak interacting massive particles (WIMPS) 
in supersymmetric models with R-parity, with particle mass $ m \sim 10-1000 $ GeV 
seriously disfavoured by galaxy observations, and finally
Warm Dark Matter (WDM), mainly sterile neutrinos with particle mass in the keV scale.

\medskip

DM particles decouple due to the universe expansion, their distribution function 
{\bf freezes out} at decoupling. The characteristic length scale after decoupling
is the {\bf free streaming scale (or Jeans' scale)}. Following the DM evolution since
ultrarelativistic decoupling by solving the linear Boltzmann-Vlasov equations
yields (see for example \cite{bdvs}),
\be\label{fs}
r_{Jeans} = 57.2 \, {\rm kpc}
\; \frac{\rm keV}{m} \; \left(\frac{100}{g_d}\right)^{\! \frac13} \; , 
\ee
where $ g_d $ equals the number of UR degrees of freedom at decoupling. 

{\vskip 0.3cm} 

DM particles can {\bf freely} propagate over distances of the order of the free streaming scale.
Therefore, structures at scales smaller or of the order of $ r_{Jeans} $ are {\bf erased}
for a given value of $ m $.

{\vskip 0.3cm} 

The observed size of the DM galaxy substructures is in the 
$ \sim 1 - 100 $ kpc scale. Therefore, eq.(\ref{fs}) indicates that $ m $ 
should be in the keV scale. That is, Warm Dark Matter particles.
This indication is confirmed by the phase-space density observations in galaxies
\cite{dvs} and further relevant evidence from galaxy observations
\cite{nos,mnrpao,tgal,dvss,satel,satel2,tikho,simuwdm,chalo,mfl,rken}.

{\vskip 0.3cm} 

For a dark matter particle decoupling at thermal equilibrium (thermal relic),
all evidences point out to a 2 keV particle. Sterile neutrinos decouple
out of thermal equilibrium with a primordial power spectrum similar
to a thermal relic but for a different particle mass \cite{nosprd,otroprd}.
More precisely, eq.(\ref{conve}) shows that a 7 keV sterile neutrino
provides a similar primordial power than a 2 keV thermal relic.

{\vskip 0.3cm} 

Therefore, WDM can be formed by 7 keV sterile neutrinos.
Notice that this result is independent of the sterile neutrino decay detection 
claimed in \cite{bulbul} of a  7 keV sterile neutrino decay. This is a further argument
in favour of 7 keV sterile neutrino WDM (see also \cite{aba}).

\bigskip

For CDM particles with $ m \sim 100$ GeV we have $ r_{Jeans} \sim 0.1 $ pc.
Hence CDM structures keep forming till scales as small as the solar system.
This result from the linear regime is confirmed  as a {\bf robust result} by
$N$-body CDM simulations. However, it has {\bf never been observed} in the sky. 

{\vskip 0.2cm} 

Adding baryons to CDM does not cure this serious problem \cite{bario}. 
There is {\bf over abundance} of small structures in CDM and in CDM+baryons
(also called the satellite problem). 

{\vskip 0.1cm} 

CDM has {\bf many serious} conflicts with observations as:
\begin{itemize}
\item{Galaxies naturally grow through merging in CDM models.
Observations show that galaxy mergers are {\bf rare} ($ <10 \% $).}
\item{Pure-disk galaxies (bulgeless) are observed whose formation 
through CDM is unexplained}.
\item{CDM predicts {\bf cusped} density profiles: $ \rho(r) \sim 1/r $ for small $ r $.
Observations show {\bf cored } profiles: $ \rho(r) $  bounded for small $ r $.
Adding by hand strong enough feedback from baryons in the CDM models
can eliminate cusps but spoils the star formation rate.}
\end{itemize}

Structures in the Universe as galaxies and cluster of galaxies
form out of the small primordial quantum fluctuations originated by inflation
just after the big-bang.

{\vskip 0.2cm} 

The linear small primordial  fluctuations grow due to gravitational (Jeans) unstabilities
and then classicalize. Structures form through non-linear gravitational evolution.
Hierarchical structure formation starts from small scales and develops to large scales.

{\vskip 0.2cm} 

$N$-body CDM simulations {\bf fail} to produce the observed structures 
for {\bf small} scales less than some kpc.

{\vskip 0.2cm} 

Both $N$-body WDM and CDM simulations yield {\bf identical and correct} structures 
for scales larger than some kpc.

{\vskip 0.2cm} 

At intermediate scales WDM give the {\bf correct abundance} of substructures \cite{simuwdm}.

{\vskip 0.2cm} 

Inside galaxy cores, below  $ \sim 100$ pc, $N$-body classical physics simulations 
are incorrect for WDM because quantum effects are important in WDM at these scales.
WDM predicts correct structures for small scales (below kpc) when its {\bf quantum} nature is
taken into account \cite{nos,mnrpao,tgal}.

{\vskip 0.2cm} 

\begin{figure}[h]
\begin{center}
\begin{turn}{-90}
\includegraphics[height=9.cm,width=6.cm]{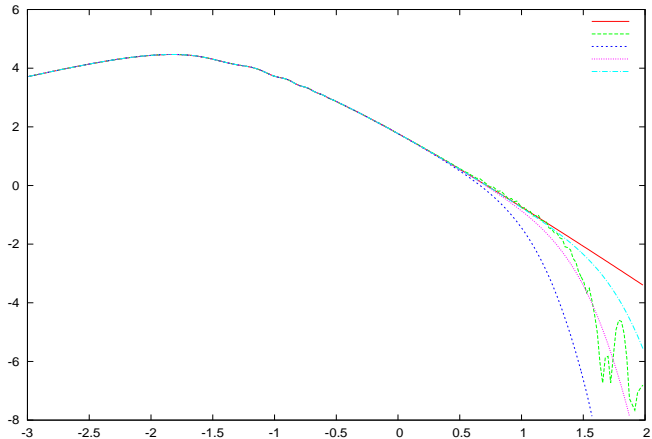}
\end{turn}
\caption{ $ \log_{10} P(k) $ vs. $ \log_{10}[k \;  {\rm Mpc}  \; h] $ 
for CDM in red, for  1  keV WDM in blue, 2 keV in violete and 4  keV  WDM in light-blue
DM particles decoupling in thermal equilibrium. 1 \, keV WDM sterile neutrinos 
decoupling out of thermal equilibrium are plotted in green. 
For WDM $ P(k) $ is cutted-off on small scales compared with CDM, that is 
$r \; \lesssim \; 100 \; ({\rm keV}/m)^{4/3} $ kpc.}
\label{potp}
\end{center}
\end{figure}

The first ingredient in structure formation is the primordial power spectrum $ P(k) $.
We plot $ P(k) $ in fig. \ref{potp} for CDM and for several examples of WDM.
CDM and WDM give identical results for the CMB fluctuations spectrum, this spectrum
corresponds to large scales $ \gtrsim 1 $ Mpc, in which WDM and CDM coincide.

{\vskip 0.2cm} 

All searches of CDM particles (wimps) look for $ m \gtrsim 1 $ GeV \cite{detecdm}.
The fact that the DM mass is in the keV scale explains why no detection 
has been reached so far.
Moreover, past, present and future reports of signals of such CDM experiments
{\bf cannot be due to DM detection} because the DM particle mass is in the keV scale.
The inconclusive signals in such experiments should be originated by phenomena
of other kinds. Notice that the supposed wimp detection signals reported in refs. \cite{detecdm}
contradict each other supporting the notion that these  signals are {\bf unrelated to any DM
detection}.

Positron excess in cosmic rays are unrelated to DM physics but to astrophysical
sources and astrophysical mechanisms and can
be explained by them \cite{BBS}.

{\vskip 0.2cm} 

Warm dark matter has been the central subject of attention 
of recent Chalonge Colloquiums whose `Highlights and Conclusions'
are online \cite{chalo}.

\section{Quantum Dark Matter physics in Galaxies}

In order to determine whether a physical system has a classical or quantum nature
one has to compare the average distance between particles $ d $ with their
de Broglie wavelength $ \lambda_{dB} $.

\medskip

The de Broglie wavelength of DM particles in a galaxy can be expressed as
\be\label{LdB}
\lambda_{dB}   = \frac{h}{m \; v } \; ,
\ee
where $ h $ stands for Planck's constant and $ v $ is the velocity dispersion,
while the average interparticle distance $ d $ can be estimated as
\be\label{dis}
d = \left( \frac{m}{\rho_h} \right)^{\! \! \frac13} \; ,
\ee
where $ \rho_h $ is the average density in the  galaxy core.
We can measure the classical or quantum character of the system by considering the ratio
$$ 
{\cal R} \equiv \frac{\lambda_{dB}}{d}
$$
For $ {\cal R} \lesssim 1 $ the system is of classical nature while for $ {\cal R} \gtrsim 1 $
it is a quantum system.

By using the phase-space density,
$$
 Q_h \equiv \frac{\rho_h}{\sigma^3}
$$
and eqs.(\ref{LdB})-(\ref{dis}), $ \cal R $ can be expressed as \cite{nos}
\be
{\cal R} = \hbar \; \frac{2 \; \pi}{\sqrt3} \; \left( \frac{Q_h}{m^4}\right)^{\! \! \frac13} \; .
\ee
Notice that $ \cal R $ as well as  $ Q_h $ are invariant under the expansion of the universe
because the lengths $ \lambda_{dB} $ and $ d $ both scale with the expansion scale factor.  $ \cal R $ and $ Q_h $
evolve by nonlinear gravitational relaxation.

Using now the observed  values of $ Q_h $ from Table \ref{pgal} yields $ \cal R $ in the range 
\be\label{quant}
 7 \times 10^{-3}  \; \left( \displaystyle \frac{\rm keV}{m}\right)^{\! \frac43}
< {\cal R} < 5 \; \left( \displaystyle \frac{\rm keV}{m}\right)^{\! \frac43}
\ee
The larger value of $ \cal R $ is for ultracompact dwarfs while the smaller value of $ \cal R $ 
is for big spirals.

\medskip

The ratio $ \cal R $ around unity clearly implies a macroscopic quantum object.
Notice that $ \cal R $ expresses solely in terms of $ Q $ and hence 
$ (\hbar^3 \; Q/m^4) $ measures how quantum or classical is the system, here, the galaxy. 
Therefore, eq.(\ref{quant}) clearly shows {\bf solely from observations} 
that compact dwarf galaxies are natural macroscopic quantum objects for WDM \cite{nos}.

\medskip

We see from eq.(\ref{quant}) that for CDM, that is for $ m\gtrsim $ GeV,
$$
{\cal R}_{CDM} \lesssim 5 \; 10^{-8}
$$
and therefore quantum effects are negligible in CDM.

\subsection{WDM Quantum pressure vs. gravitational pressure in compact galaxies}\label{qup}

For an order--of--magnitude estimate, let us consider a halo of mass $ M $ and radius 
$ R $ of fermionic matter. Each fermion can be considered inside
a  cell of size $ \Delta x \sim 1 / n^{\frac13} $ and therefore has a momentum
$$
p \sim \frac{\hbar}{\Delta x} \sim \hbar \; n^{\frac13} \; .
$$
The associated quantum pressure $ P_q $ (flux of the momentum) has the value 
\be\label{presq}
P_q = n \; v \; p \sim \hbar \; v \; n^{\frac43} = \frac{\hbar^2}{m} \; n^{\frac53} \; .
\ee
where $ v $ is the mean velocity given by
$$
v = \frac{p}{m} = \frac{\hbar}{m} \; n^{\frac13} \; .
$$ 
The number density can be estimated as
$$
n = \frac{M}{\frac43 \; \pi \; R^3 \; m} \; ,
$$
and we obtain from eq.(\ref{presq}) the quantum pressure
\be\label{pqu}
 P_q = \frac{\hbar^2}{m \; R^5} \; \left(\frac{3 \; M}{4 \; \pi \; m}\right)^{\! \! \frac53} \; .
\ee
On the other hand, as is well known, galaxy formation as all structure formation in the Universe
is driven by gravitational physics.
The system will be in dynamical equilibrium if the quantum pressure is balanced by
the gravitational pressure
\be\label{pgr}
P_G = {\rm gravitational~ force}/{\rm area} = \frac{G \; M^2}{R^2} \times \frac1{4 \, \pi \; R^2}
\ee
Equating $  P_q = P_G $ from eqs.(\ref{pgr})-(\ref{pqu})
yields the following expressions for the size $ R $ and the velocity $ v $ 
in terms of the mass $ M $ of the system and the mass $ m $ of the particles \cite{nos}:
\bea\label{estM}
&& R = \frac{3^\frac53}{(4 \; \pi)^\frac23} \;
\frac{\hbar^2}{G \; m^\frac83 \; M^\frac13} = 7.8 \ldots {\rm pc} \;
\left(\frac{10^4 \;  M_\odot}{ M}\right)^{\! \! \frac13} \; 
\left(\frac{2 \; \rm keV}{m}\right)^{\! \! \frac83} \; , \\ \cr
&& v = \sqrt3 \; \sigma = \sqrt3 \; 
\left(\frac{4 \, \pi}{81}\right)^{\! \! \frac13} \; \frac{G}{\hbar} \;  m^\frac43 \; M^\frac23=
4.64 \ldots \frac{\rm km}{\rm s} \; \left(\frac{m}{2 \; \rm keV}\right)^{\! \! \frac43} \; 
\left(\frac{M}{10^4 \;  M_\odot}\right)^{\! \! \frac23} \;   .
\eea
Notice that the values of $ M , \;  R $ and $ v $ are consistent with the observed values of
ultracompact dwarf galaxies. Namely, for $ M $ of the order $ 10^4 \;  M_\odot $
(which is a typical mass value for an ultracompact dwarf galaxies), $ R $ and  $ v $ 
give the correct order of magnitude for the size and velocity dispersion of dwarf galaxies 
as displayed in Table \ref{pgal}, for WDM particle mass in the keV scale.
These estimates are in agreement with the precise Thomas--Fermi results in the degenerate limit \cite{tgal}.

These results back the idea that dwarf galaxies are supported by the
fermionic {\it WDM quantum pressure} eq.(\ref{pqu}) \cite{nos}.

\begin{table}
\begin{tabular}{|c|c|c|c|c|c|} \hline  
 & & & & & \\
 Galaxy  & $ \displaystyle \frac{r_h}{\rm pc} $ & $  \displaystyle \frac{v}{\frac{\rm km}{\rm s}} $
& $ \displaystyle  \frac{\hbar^{\frac32} \;\sqrt{Q_h}}{({\rm keV})^2} $ & 
$ \rho(0)/\displaystyle \frac{M_\odot}{({\rm pc})^3} $ & $ \displaystyle \frac{M_h}{10^6 \; M_\odot} $
\\ & & & & & \\ \hline 
Willman 1 & 19 & $ 4 $ & $ 0.85 $ & $ 6.3 $ & $ 0.029 $
\\ \hline  
 Segue 1 & 48 & $ 4 $ & $ 1.3 $ & $ 2.5 $ & $ 1.93 $ \\ \hline  
  Leo IV & 400 & $ 3.3 $ & $ 0.2 $ & $ .19 $ & $ 200 $ \\ \hline  
Canis Venatici II & 245 & $ 4.6 $ & $ 0.2 $   & $ 0.49 $ & $ 4.8 $
\\ \hline  
Coma-Berenices & 123 & $ 4.6 $  & $ 0.42 $   & $ 2.09 $  & $ 0.14 $
\\ \hline  
 Leo II & 320 & $ 6.6 $ & $ 0.093 $  & $ 0.34 $ & $ 36.6 $
\\  \hline  
 Leo T & 170 & $ 7.8 $ &  $ 0.12 $  & $ 0.79 $ & $ 12.9 $
\\ \hline  
 Hercules & 387 & $ 5.1 $ &  $ 0.078 $  & $ 0.1 $ & $ 25.1 $
\\ \hline  
 Carina & 424 & $ 6.4 $ & $ 0.075 $  & $ 0.15 $ & $ 32.2 $
\\ \hline 
 Ursa Major I & 504 & 7.6  &  $ 0.066 $  & $ 0.25 $ & $ 33.2 $
\\ \hline  
 Draco & 305 & $ 10.1 $ &  $ 0.06 $  & $ 0.5 $ & $ 26.5 $
\\ \hline  
 Leo I & 518  & $ 9 $ &  $ 0.048 $  & $ 0.22 $ & $ 96 $
\\ \hline  
 Sculptor & 480  & $ 9 $ & $ 0.05 $  & $ 0.25  $ & $ 78.8 $
\\ \hline 
 Bo\"otes I & 362 & $ 9 $ & $ 0.058 $  & $ 0.38 $ & $ 43.2 $
\\ \hline  
 Canis Venatici I & 1220  & $ 7.6 $ & $ 0.037 $ & $ 0.08 $ & $ 344 $
\\ \hline  
Sextans & 1290 & $ 7.1 $ & $ 0.021 $ & $ 0.02 $ & $ 116 $
\\ \hline 
 Ursa Minor & 750 & $ 11.5 $ & $ 0.028 $  & $ 0.16 $ & $ 193 $
\\ \hline  
 Fornax  & 1730 & $ 10.7 $ & $ 0.016 $  & $ 0.053  $ & $ 1750 $
\\  \hline  
 NGC 185  & 450 & $ 31 $ & $ 0.033 $ & $ 4.09 $ & $ 975 $
\\ \hline  
 NGC 855  & 1063 & $ 58 $ & $ 0.01 $ & $ 2.64 $ & $ 8340 $
\\ \hline  
  Small Spiral  & 5100  & $ 40.7 $ & $ 0.0018 $ & $ 0.029 $ & $ 6900 $
\\ \hline  
NGC 4478 & 1890 & $ 147 $ & $ 0.003 $ & $ 3.7 $ & $ 6.55 \times 10^4 $
\\ \hline  
 Medium Spiral & $ 1.9 \times 10^4 $ & $ 76.2 $ & $ 3.7 \times 10^{-4} $ & $ 0.0076 $ & $ 1.01 \times 10^5 $
\\ \hline  
 NGC 731 & 6160 & $ 163 $ & $ 9.27 \times 10^{-4} $ & $ 0.47 $ & $ 2.87 \times 10^5 $
\\ \hline 
 NGC 3853   & 5220 & $ 198 $ & $ 8.8 \times 10^{-4} $ & $ 0.77 $  
& $ 2.87 \times 10^5 $ \\ \hline 
NGC 499  & 7700 &  $ 274 $ & $ 5.9 \times 10^{-4} $ & 
$ 0.91 $ & $ 1.09 \times 10^6 $ \\   \hline 
Large Spiral & $ 5.9 \times 10^4 $ & $ 125 $ & $ 0.96 \times 10^{-4} $ & $ 2.3 \times 10^{-3} $ & 
$ 1. \times 10^6 $ \\ \hline  
\end{tabular}
\caption{Observed values $ r_h $, velocity dispersion $ v, \;  \sqrt{Q_h}, \; \rho(0)$ 
and $ M_h $ covering from ultracompact galaxies to large spiral galaxies
from refs.\cite{wp,sltg,dvss,gil,jdsmg,datos}. The phase space density is larger
for smaller galaxies, both in mass and size.
Notice that the phase space density is obtained
from the stars velocity dispersion which is expected to be smaller than the DM  velocity dispersion.
Therefore, the reported $ Q_h $ are in fact upper bounds to the true values \cite{jdsmg}.}
\label{pgal}
\end{table}

\section{Quantum  fermionic WDM gives the correct galaxy properties and cored galaxy profiles}\label{tf}

We treat here the self-gravitating fermionic DM in the Thomas-Fermi approximation.
In this approach, the central quantity to derive is the DM chemical potential $ \mu(r) $,
the chemical potential being the free energy per particle \cite{ll}.
We consider a single DM halo in the late stages of structure formation when DM
particles composing it are non--relativistic and their phase--space distribution
function $ f(t, \br,\bp) $ is relaxing to a time--independent form, at least for
$\br$ not too far from the halo center. In the Thomas--Fermi approach such a
time--independent form is taken to be a energy distribution function $f(E)$ of
the conserved single--particle energy $E = p^2/(2m) - \mu $, where $m$ is the
mass of the DM particle and $\mu$ is the chemical potential
\be \label{potq}
  \mu(\br) =  \mu_0 - m \, \phi(\br) 
\ee
with $\phi(\br)$ the gravitational potential and  $ \mu_0 $ some constant.
We consider the spherical symmetric case. 

Here, the Poisson equation for $ \phi(r) $ is a nonlinear and selfconsistent equation
\be \label{pois}
  \frac{d^2 \mu}{dr^2} + \frac2{r} \; \frac{d \mu}{dr} = - 4\pi \, G \, m \, \rho(r)\; , 
\ee
where the mass density $ \rho(r) $ is a function of $ \mu(r) $ and
$ G $ is Newton's constant. $ \rho(r) $ is expressed here as a function of $ \mu(r) $ through the
standard integral of the DM phase--space distribution function over the momentum
for Dirac fermions as
\be \label{den}
  \rho(r) = \frac{m}{\pi^2 \, \hbar^3} \int_0^{\infty} dp\;p^2 
  \; f\left(\displaystyle \frac{p^2}{2m}-\mu(r)\right)\; , 
\ee
Another standard integral of the DM phase--space distribution function is the pressure
\be \label{P}
  P(r) = \frac{1}{3\pi^2 \,m\,\hbar^3} \int_0^{\infty} dp\;p^4 
  \,f\left(\displaystyle \frac{p^2}{2m}-\mu(r)\right) \; .
\ee
From $ \rho(r) $ and $ P(r) $ other quantities of interest, such as the velocity dispersion
$ \sigma(r) $ and the phase--space density $ Q(r) $ can be determined as
\be \label{sigQ}
  \sigma^2(r) = \frac{P(r)}{\rho(r)} \quad,\qquad Q(r) = \frac{\rho(r)}{\sigma^3(r)}  \; .
\ee
We see that $\mu(r)$ fully characterizes the fermionic DM halo in this
Thomas--Fermi framework. The chemical potential is monotonically decreasing in $ r $ 
since eq.~(\ref{pois}) implies
\be\label{dmu}
\frac{d\mu}{dr} = -\frac{G\,m\,M(r)}{r^2} \quad,\qquad  
  M(r) = 4\pi \int_0^r dr'\, r'^2 \, \rho(r') \; .
\ee
Moreover, the fermionic DM mass density $ \rho$ is bounded at the origin due to the Pauli principle \cite{nos},
and therefore the proper boundary condition at the origin is
\be
  \frac{d \mu}{dr}(0) = 0 \; .
\ee
Eqs.(\ref{pois}) and (\ref{den}) provide an ordinary nonlinear
differential equation that determines selfconsistently the chemical potential $ \mu(r) $ and
constitutes the Thomas--Fermi semi-classical approach
\cite{nos,mnrpao,tgal} (see also ref. \cite{peter}).
We obtain a family of solutions 
parametrized by the value of  $ \mu_0 \equiv \mu(0) $ \cite{nos}.

\medskip

In this semi-classical framework the stationary energy distribution function $
f(E) $ must be assigned beforehand. In a full--fledged treatment one would
solve the cosmological WDM evolution since decoupling till today, including the quantum
WDM dynamics in the evolution which become important in the non-linear stage and close
enough to the origin. 

\medskip

We integrate the Thomas-Fermi nonlinear differential
equations (\ref{pois})-(\ref{den}) from $ r = 0 $ till the 
boundary $ r = R = R _{200} \sim R_{vir} $ defined as the radius where the 
mass density equals $ 200 $ times the mean DM density \cite{nos}.

We define the core size $ r_h $ of the halo by analogy with the Burkert density profile as
\be\label{onequarter}
  \frac{\rho(r_h)}{\rho_0} = \frac14 \quad , \quad  r_h = l_0 \; \xi_h \; .
\ee
where $ \rho_0 \equiv \rho(0) $ and $ l_0 $ is the characteristic length that emerges from 
the dynamical equations (\ref{pois})-(\ref{den}):
\be\label{varsd2}
l_0 \equiv  \frac{\hbar}{\sqrt{8\,G}} \left(\frac{9\pi}{m^8\,\rho_0}\right)^{\! \! \frac16} 
  = R_0 \; \left(\frac{{\rm keV}}{m}\right)^{\! \! \frac43}  \; 
  \left(\rho_0 \; \frac{{\rm pc}^3}{M_\odot}\right)^{\! \! -\frac16} 
  \;,\qquad R_0 = 18.71 \; \rm pc  \; .
\ee
To explicitly solve eqs.(\ref{pois})-(\ref{den}) we need to specify the distribution function
$ \Psi(E/E_0) $. But many important properties of the Thomas--Fermi semi-classical
approximation do not depend on the detailed form of the distribution function
$ \Psi(E/E_0) $. Indeed, a generic
feature of a physically sensible one--parameter form $ \Psi(E/E_0) $ is that it should
describe degenerate fermions for $ E_0 \to 0 $. That is, $ \Psi(E/E_0) $ should behave as
the step function $ \theta(-E) $ in such limit. In the opposite limit, $ \mu/E_0 \to -\infty $,
$ \Psi(E/E_0) $ describes classical particles, namely a Boltzmann distribution.
As an example of distribution function, we consider the Fermi--Dirac distribution 
\be\label{FD}
  \Psi_{\rm FD}(E/E_0) = \frac1{e^{E/E_0} + 1} \; .
\ee
We define the dimensionless chemical potential $ \nu(r) $ as  
$$ 
\nu(r) \equiv \mu(r)/E_0 \quad {\rm and} \quad   \nu_0 \equiv \mu(0)/E_0 \quad .
$$
Positive values of the chemical potential at the origin $ \nu_0 > 1 $ 
correspond to the fermions in the quantum regime, and oppositely, $ \nu_0 < -1 $ gives 
the diluted regime which is the classical regime. In this classical regime the Thomas-Fermi equations
(\ref{pois})-(\ref{den}) become exactly the equations for a self-gravitating Boltzmann gas.

\medskip

\begin{figure}
\begin{turn}{-90}
\psfrag{"Kperfu10.dat"}{$ {\hat M}_h= 7 \; 10^{11}  \; M_\odot $}
\psfrag{"Kperfu11.dat"}{$ {\hat M}_h= 6.2 \; 10^8  \; M_\odot $}
\psfrag{"Kperfu12.dat"}{$ {\hat M}_h= 1.3 \; 10^8 \; M_\odot  $}
\psfrag{"Kperfu13.dat"}{$ {\hat M}_h= 2.5 \; 10^7 \; M_\odot  $}
\psfrag{"Kperfu14.dat"}{$ {\hat M}_h= 5.1 \; 10^6 \; M_\odot  $}
\psfrag{"Kperfu15.dat"}{$ {\hat M}_h= 1.1 \; 10^6 \; M_\odot  $}
\psfrag{"Kperfu16.dat"}{$ {\hat M}_h= 2.2 \; 10^5 \; M_\odot  $}
\psfrag{"Kperfu30.dat"}{$ {\hat M}_h= 1.6 \; 10^5 \; M_\odot  $}
\includegraphics[height=12.cm,width=7.cm]{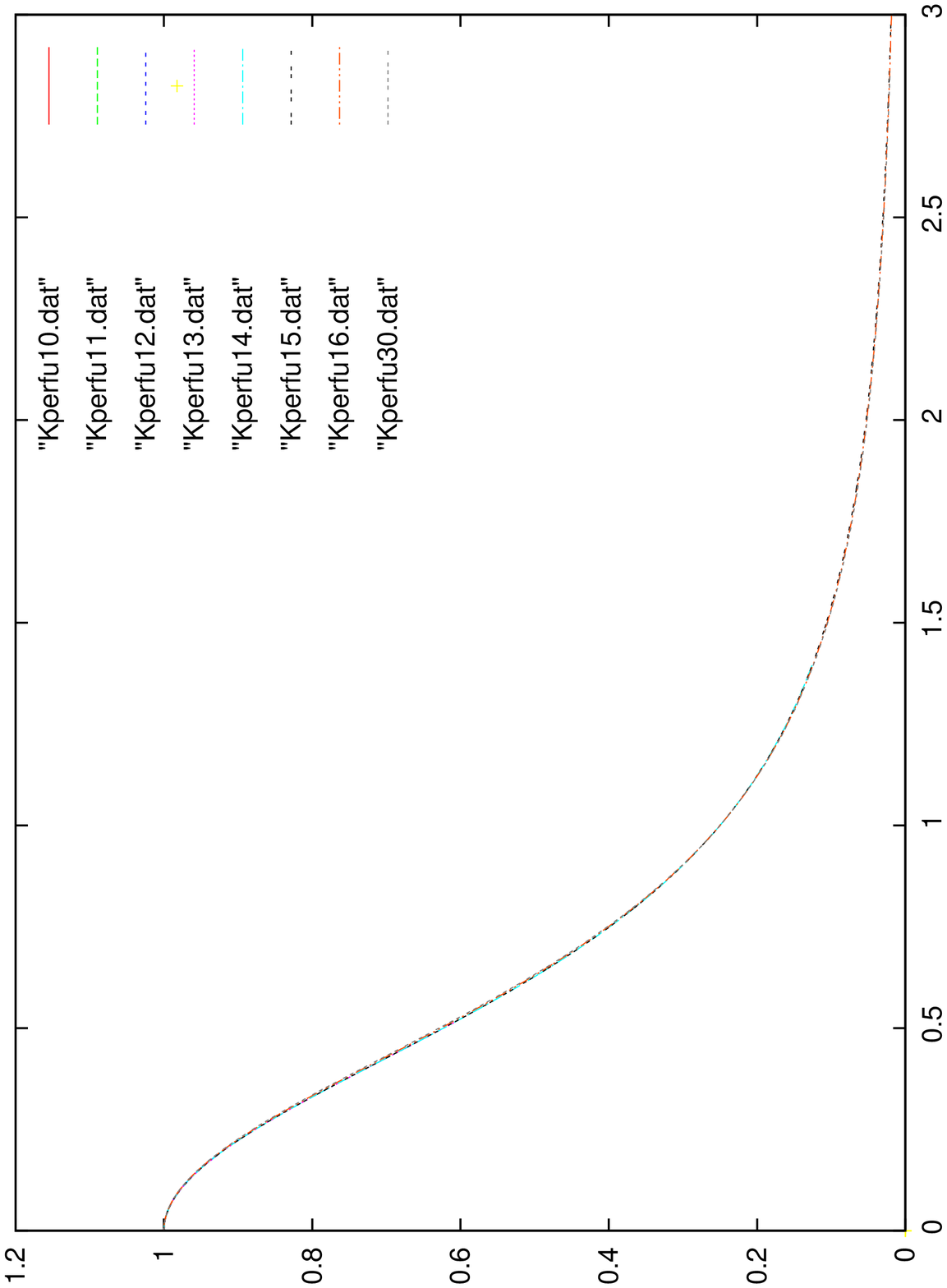}
\psfrag{"Rperfu11.dat"}{universal diluted profile}
\psfrag{"Rperfu17.dat"}{$ {\hat M}_h= 3.9 \; 10^4 \; M_\odot  $}
\psfrag{"Rperfu18.dat"}{$ {\hat M}_h= 3.4 \; 10^4 \; M_\odot  $}
\psfrag{"Rperfu19.dat"}{$ {\hat M}_h= 3.3 \; 10^4 \; M_\odot  $}
\psfrag{"Rperfu20.dat"}{$ {\hat M}_h= 3.2 \; 10^4 \; M_\odot  $}
\psfrag{"Rperfudeg.dat"}{degenerate limit}
\includegraphics[height=12.cm,width=7.cm]{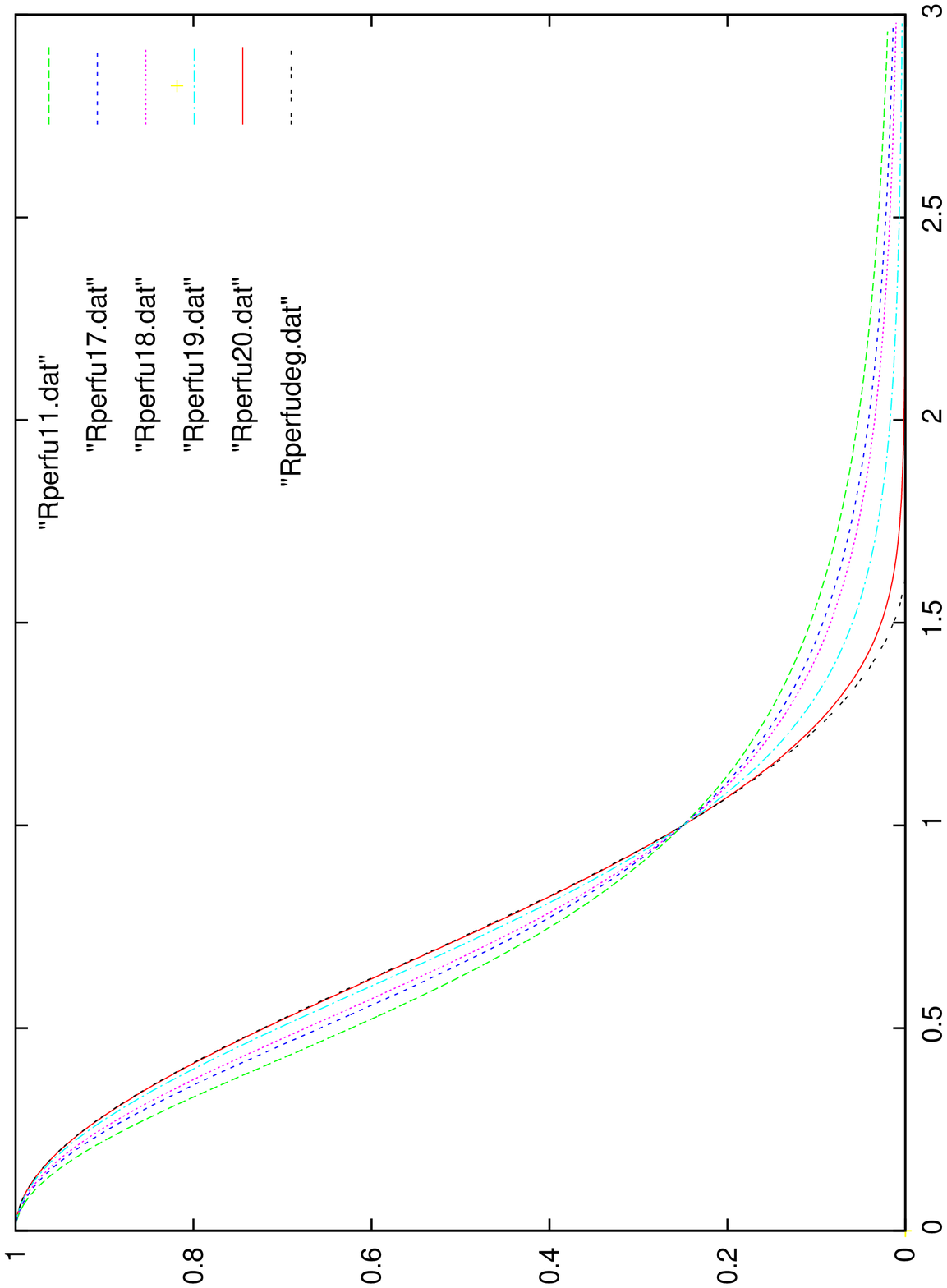}
\end{turn}
\caption{Normalized density profiles $ \rho(r)/\rho(0) $ as functions of $ r/r_h $.
We display in the upper panel the profiles for galaxy masses in the diluted regime
$  1.4 \; 10^5 \; M_\odot < {\hat M}_h < 7.5 \; 10^{11}\; M_\odot , \; -1.5 > \nu_0 > -20.78 $ which {\bf all} provide 
the {\bf same universal} density profile. We display in the lower panel the profiles for
galaxy masses $  M_h^{min} = 30999 \; \left(2 \, {\rm keV}/m\right)^{\! \! \frac{16}5} \; M_\odot
 \leq {\hat M}_h <  3.9 \; 10^4 \; M_\odot, \; 1 < \nu_0 < \infty $ which are near the quantum 
degenerate regime and exhibit shrinking density profiles for decreasing galaxy mass.
For comparison, we also plot in the lower figure the universal profile in the diluted regime.}
\label{perfus}
\end{figure}

Normalizing the density profiles as $ \rho(r)/\rho(0) $ 
and plotting them as functions of $ r/r_h $ produce normalized profiles which 
are {\bf universal} functions of $ x \equiv r/r_h $ in the diluted regime as shown in fig. \ref{perfus}.
This universality is valid for {\bf all} galaxy masses $ {\hat M}_h > 10^5  \; M_\odot $ \cite{tgal}.
The obtained fermion profiles are always cored. 

{\vskip 0.1cm} 

Our theoretical density profiles and rotation curves obtained from the 
Thomas-Fermi equations remarkably agree with observations for $ r \lesssim r_h $, for all 
galaxies in the diluted regime \cite{mnrpao}. This indicates that WDM is thermalized in the internal regions 
$ r \lesssim r_h $ of galaxies \cite{tgal}.

{\vskip 0.2cm} 

For galaxy masses $ {\hat M}_h < 10^5  \; M_\odot $, near the quantum degenerate regime, the normalized density profiles 
$ \rho(r)/\rho(0) $ are not anymore universal and depend on the galaxy mass.

{\vskip 0.1cm} 

As we can see in fig. \ref{perfus} the density profile shape changes fastly
when the galaxy mass decreases only by a factor seven from $ {\hat M}_h = 1.4 \; 10^5  \; M_\odot $ to the minimal galaxy
mass $ {\hat M}_{h,min} = 3.10 \; 10^4 \; M_\odot $. In this narrow range of galaxy masses the density profiles 
shrink from the universal profile till the degenerate profile as shown in fig. \ref{perfus}.
Namely, these dwarf galaxies are more compact than the larger diluted galaxies.

\medskip

\begin{figure}
\begin{turn}{-90}
\psfrag{"perfv11.dat"}{$ {\hat M}_h= 7.1 \; 10^{11 \; M_\odot } $}
\psfrag{"perfv12.dat"}{$ {\hat M}_h= 6.3 \; 10^9 \; M_\odot  $}
\psfrag{"perfv13.dat"}{$ {\hat M}_h= 1.2 \; 10^8 \; M_\odot  $}
\psfrag{"perfv14.dat"}{$ {\hat M}_h= 2.3 \; 10^6 \; M_\odot  $}
\psfrag{"perfv15.dat"}{$ {\hat M}_h= 2.2 \; 10^5 \; M_\odot  $}
\psfrag{"perfv16.dat"}{$ {\hat M}_h= 6.1 \; 10^4 \; M_\odot  $}
\psfrag{"perfv17.dat"}{$ {\hat M}_h= 4.0 \; 10^4 \; M_\odot  $}
\psfrag{"perfv18.dat"}{$ {\hat M}_h= 3.0 \; 10^4 \; M_\odot  $}
\psfrag{"perfv19.dat"}{$ {\hat M}_h= 2.2 \; 10^4 \; M_\odot  $}
\psfrag{"perfv20.dat"}{$ {\hat M}_h= 2.0 \; 10^4 \; M_\odot  $}
\psfrag{"perfvdeg.dat"}{degenerate limit}
\includegraphics[height=12.cm,width=10.cm]{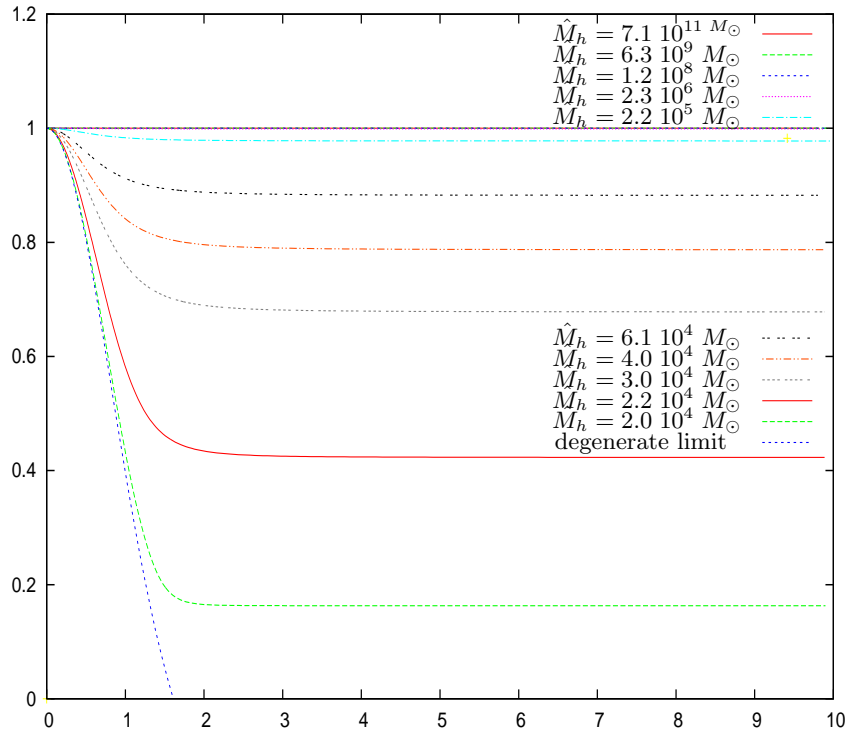}
\end{turn}
\caption{Normalized velocity dispersion profiles $ \sigma^2(r)/\sigma^2(0) $ as functions of $ x=r/r_h $.
All velocity profiles in the diluted regime for galaxy masses $ {\hat M}_h > 2.3 \; 10^6 \; M_\odot, 
\; \nu_0 < -5 $ fall into
the same {\bf constant universal} profile corresponding to a perfect but inhomogeneous self-gravitating WDM
gas describing large and diluted galaxies. 
The velocity profiles for smaller galaxy masses $ 1.6 \; 10^6 \; M_\odot> {\hat M}_h > {\hat M}_{h,min} 
= 3.10 \; 10^4 \; M_\odot $ 
do depend on $ x $ and yield decreasing velocity dispersions for decreasing galaxy masses, accounting for the quantum
fermionic effects which become important in this range of galaxy masses (WDM compact dwarf galaxies). }
\label{perfv}
\end{figure}

We display in fig. \ref{perfv} the normalized velocity dispersion profiles $ \sigma^2(r)/\sigma^2(0) $
as functions of $ x = r/r_h $. Again, we see that these profiles are {\bf universal
and constant}, i. e.  independent of the galaxy mass in the diluted regime for 
$ M_h > 2.3 \; 10^6 \; M_\odot , \; \nu_0 < -5 , \; T_0 > 0.017 $ K. The constancy of 
$ \sigma^2(r) = \sigma^2(0) $ in the diluted regime
implies that the equation of state is that of a perfect but inhomogeneous WDM gas \cite{tgal}
\be\label{sigdil}
P(r) = \frac13 <v^2>(r) \; \rho(r) = \sigma^2(r) \; \rho(r) \quad , \quad 
\sigma^2(r) = \sigma^2(0) = \frac{T_0}{m} \; ,
\ee
WDM diluted galaxies exhibit a perfect gas equation of state 
where both the pressure $ P(r) $ and the density $ \rho(r) $ depend on the coordinates.

{\vskip 0.1cm} 

For smaller galaxy masses $ 1.6 \; 10^6 \; M_\odot > {\hat M}_h > {\hat M}_{h,min} $,
the velocity profiles do depend on $ r $ and yield decreasing velocity dispersions for decreasing galaxy masses.
Namely, the deviation from the universal curves appears for $ {\hat M}_h < 10^6  \; M_\odot  $
and we see that it precisely arises from the quantum fermionic effects which become important
in such range of galaxy masses.

{\vskip 0.1cm} 

The sizes of the cores $ r_h $ defined by eq.(\ref{onequarter}) 
are in agreement with the observations, from the compact galaxies where $ r_h \sim 35 $ pc till
the spiral and elliptical galaxies where $ r_h \sim 0.2 - 60 $ kpc. The larger and positive is 
$ \nu_0 $, the smaller is the core. The minimal core size arises in
the degenerate case  $ \nu_0 \to +\infty $ (compact dwarf galaxies).

\begin{figure}
\begin{turn}{-90}
\psfrag{"Fvcirc1.dat"}{$ M_h = 5.1 \; 10^9 \; M_{\odot} $: Theory}
\psfrag{"vcsalu1.dat"}{Observational}
\psfrag{"Fvcirc2.dat"}{$ M_h = 8.4 \; 10^9\; M_{\odot} $: Theory}
\psfrag{"vcsalu2.dat"}{Observational}
\psfrag{"Fvcirc3.dat"}{$ M_h = 1.4 \; 10^{10}\; M_{\odot} $: Theory}
\psfrag{"vcsalu3.dat"}{Observational}
\psfrag{"Fvcirc4.dat"}{$ M_h = 2.3 \; 10^{10}\; M_{\odot} $: Theory}
\psfrag{"vcsalu4.dat"}{Observational}
\psfrag{"Fvcirc5.dat"}{$ M_h = 3.8 \; 10^{10} \; M_{\odot} $: Theory}
\psfrag{"vcsalu5.dat"}{Observational}
\includegraphics[height=12.cm,width=10.cm]{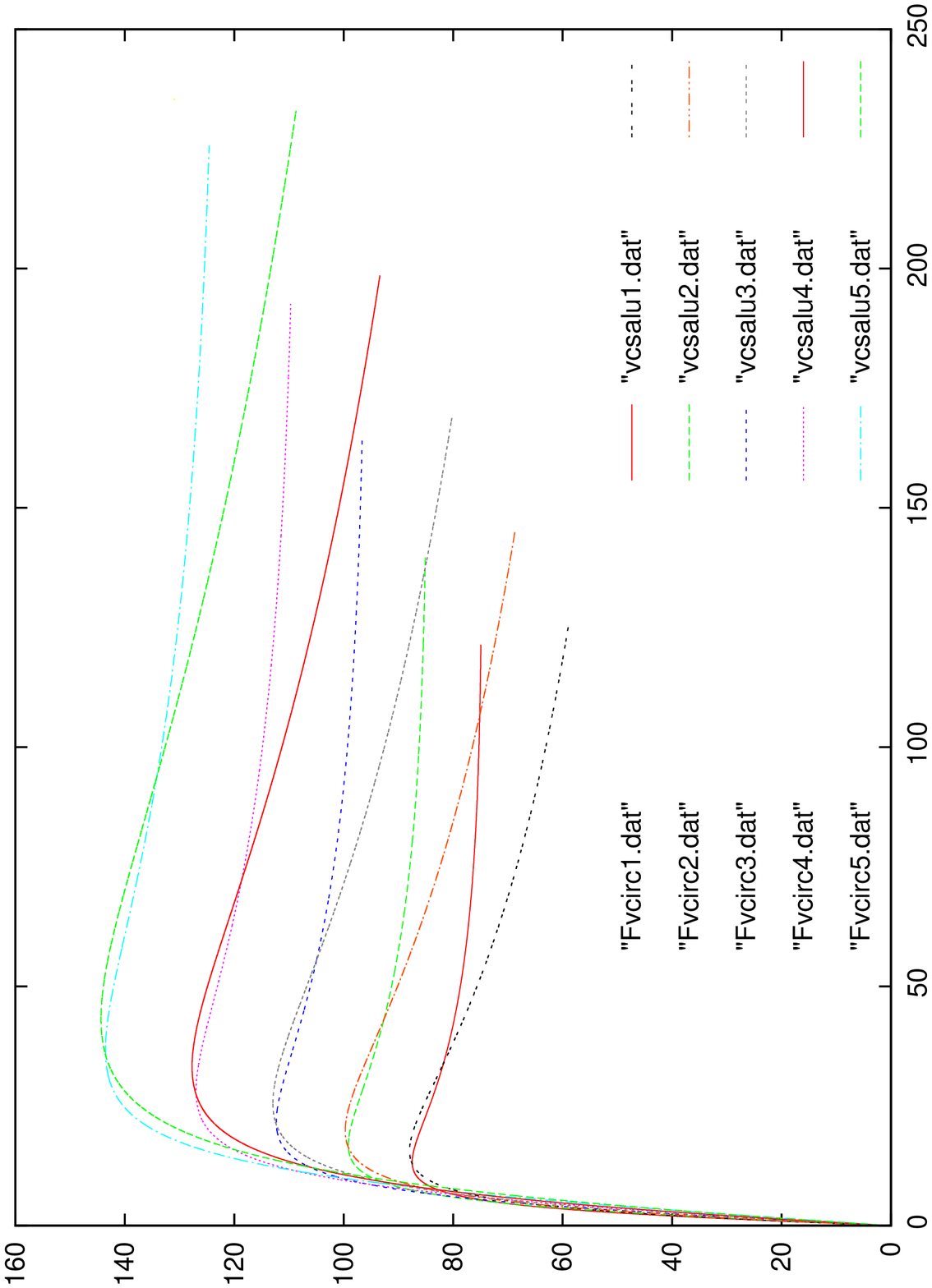}
\psfrag{"Fvcirc6.dat"}{$ M_h = 6.4 \; 10^{10} \; M_{\odot} $: Theory}
\psfrag{"vcsalu6.dat"}{Observational}
\psfrag{"Fvcirc7.dat"}{$ M_h = 1.1 \; 10^{11}\; M_{\odot} $: Theory}
\psfrag{"vcsalu7.dat"}{Observational}
\psfrag{"Fvcirc8.dat"}{$ M_h = 1.8 \; 10^{11}\; M_{\odot} $: Theory}
\psfrag{"vcsalu8.dat"}{Observational}
\psfrag{"Fvcirc9.dat"}{$ M_h = 3.0 \; 10^{11}\; M_{\odot} $: Theory}
\psfrag{"vcsalu9.dat"}{Observational}
\psfrag{"Fvcirc10.dat"}{$ M_h = 5.2 \; 10^{11} \; M_{\odot} $: Theory}
\psfrag{"vcsalu10.dat"}{Observational}
\includegraphics[height=12.cm,width=10.cm]{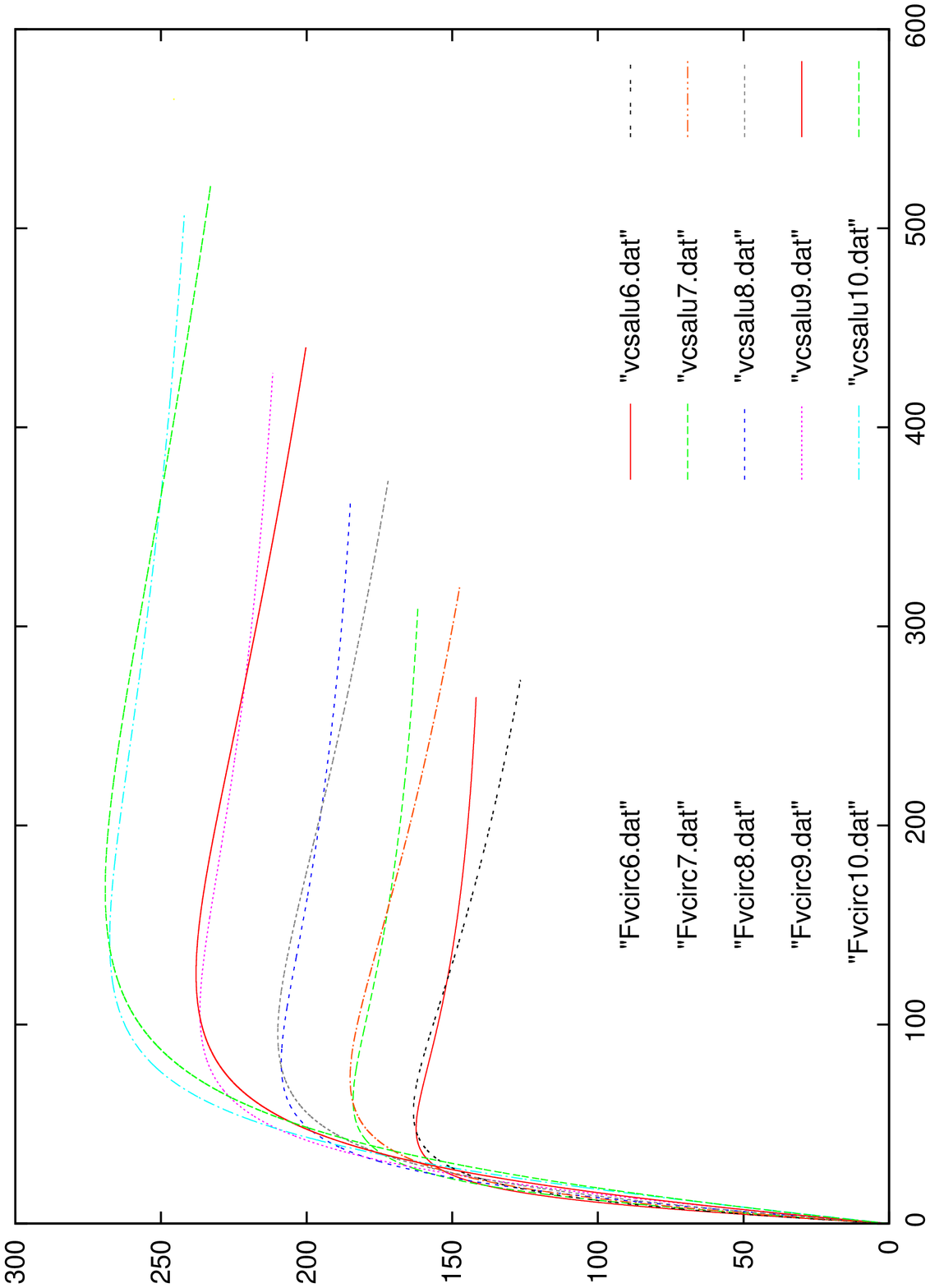}
\end{turn}
\caption{The velocity rotation curves $ v_c(r) $ in km/s versus $ r $ in kpc
for ten different independent galaxy masses $ M_h $ going from $ 5.13 \;  10^9 \; M_{\odot} $
till $ 5.15 \; 10^{11} \; M_{\odot} $. For each galaxy mass  $ M_h $,
we show the two curves: the theoretical Thomas-Fermi curve and the observational
Burkert curve. The Thomas-Fermi curves reproduce remarkably well the observational
curves for $ r \lesssim r_h $ \cite{mnrpao}. 
We plot $ v_c(r) $ for $ 0 < r < r_{vir} , \; r_{vir} $
being the virial radius of the galaxy.}
\label{vcirc}
\end{figure}

\begin{figure}
\begin{turn}{-90}
\psfrag{"RLMLrhmu120.dat"}{observed values}
\psfrag{"Mr0.dat"}{theory curve}
\includegraphics[height=12.cm,width=10.cm]{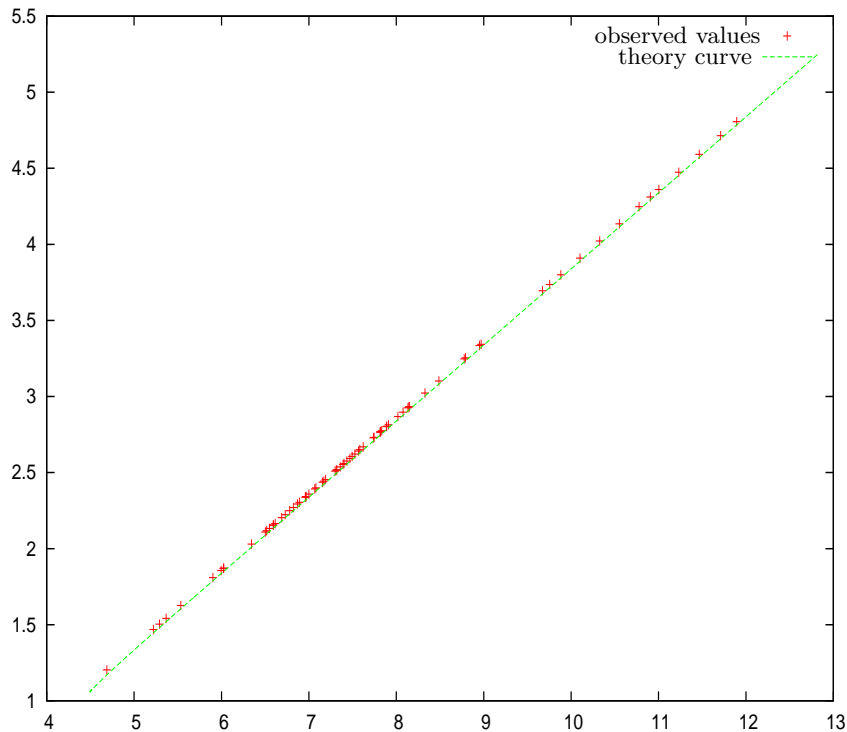}
\end{turn}
\caption{The ordinary logarithm of $ {\hat r}_h = \frac{r_h}{\rm pc} \; \left(\frac{\Sigma_0 \;  
{\rm pc}^2}{120 \; M_\odot}\right)^{\! \! \frac15} $ vs. the ordinary logarithm of 
$ {\hat M}_h = \frac{M_h}{M_\odot} \; \left(\frac{120 \; M_\odot}{\Sigma_0 \;  
{\rm pc}^2}\right)^{\! \! \frac35} $. We see that $ r_h $ follows with precision 
the square-root of $ M_h $ as in the dilute regime of the Thomas-Fermi equations \cite{tgal}. 
The data for $ M_h $ and $ r_h $ are taken from Table 1 in \cite{nos}, 
from \cite{mcc} and from \cite{sltg} and they are satisfactorily reproduced by the theoretical
Thomas-Fermi curve.}
\label{halo}
\end{figure}

\begin{figure}
\begin{turn}{-90}
\psfrag{"Cmqv.dat"}{Thomas-Fermi $ \log_{10}  Q_{TF}/{\rm keV}^4 $}
\psfrag{"Rqn.dat"}{$ Q_{Bur} $ Galaxy Data 1}
\psfrag{"qawmcc.dat"}{$ Q_{Bur} $ Galaxy Data 2}
\psfrag{"qpaolo.dat"}{$ Q_{Bur} $ Galaxy Data 3}
\includegraphics[height=12.cm,width=10.cm]{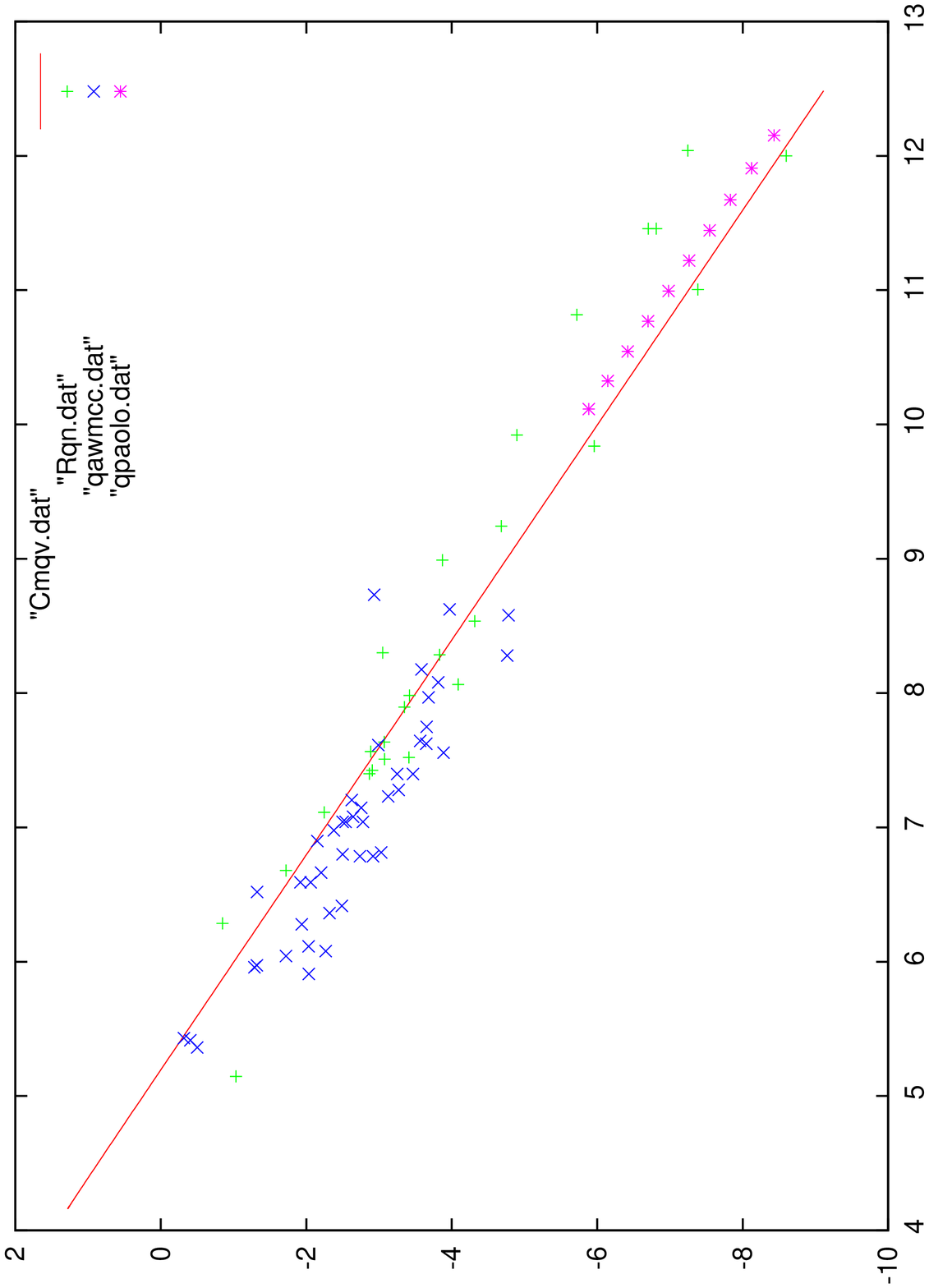}
\end{turn}
\caption{The theoretical Thomas-Fermi $ \log_{10}  Q_{TF}/{\rm keV}^4 $ phase--space density
vs. the halo mass $ \log_{10} {\hat M}_h $. The theoretical curve $ Q_{TF} $ is obtained
from the Thomas-Fermi expression \cite{mnrpao}.
The data for $ Q_{Bur} $ have been obtained from circular velocities.
Galaxy Data 1 refers to data from Table 1 in \cite{nos}, 
Galaxy Data 2 refers to data from \cite{mcc} and Galaxy Data 3 refers to data from 
\cite{sltg}.}
\label{q}
\end{figure}

We plot in fig. \ref{q} the ordinary logarithm of the theoretical Thomas-Fermi phase-space density
$ \log_{10} Q_{TF}/{\rm keV}^4 $ vs. the ordinary logarithm of $ {\hat M}_h $
and the observational values of $ \log_{10} Q_{Bur}/{\rm keV}^4 $. 
We see that the theoretical phase-space density $ Q_{TF} $
reproduces very well the observational data \cite{mnrpao}.

\medskip

We {\bf derive} the general equation of 
state for galaxies, i. e., the relation between pressure and density, and provide
its analytic expression \cite{tgal}. Two regimes clearly show up: (i) Large diluted galaxies for $ M_h \gtrsim 2.3 \;
10^6 \; M_\odot $ and effective temperatures $ T_0 > 0.017 $ K described by 
the classical selfgravitating WDM Boltzman gas with an inhomogeneous perfect gas equation of state,
and (ii) Compact dwarf galaxies for  $ 1.6 \; 10^6 \; M_\odot \gtrsim M_h \gtrsim M_{h,min} \simeq 
3.10 \; 10^4 \; \left(2 \, {\rm keV}/m\right)^{\! \! \frac{16}5} \; M_\odot, 
\; T_0 < 0.011 $ K described by the quantum fermionic WDM regime with a steeper equation of 
state close to the degenerate state. In particular, the $ T_0 = 0 $ degenerate or extreme quantum limit yields 
the most compact and smallest galaxy. All magnitudes in the diluted regime turn to exhibit square 
root of $ M_h $ {\bf scaling} laws and are {\bf universal} functions of $ r/r_h $ 
reflecting the WDM perfect gas behaviour in this regime. These theoretical results contrasted to 
robust and independent sets of galaxy data 
remarkably reproduce the observations. For the small galaxies, $ 10^6 \gtrsim M_h \geq M_{h,min} $,
the equation of state is galaxy mass dependent 
and the density and velocity profiles are not anymore universal,
accounting to the quantum physics of the self-gravitating WDM fermions in the compact regime (near, but not at, 
the degenerate state). It would be extremely interesting to dispose of dwarf galaxy 
observations which could check these quantum effects. 

\medskip 

We find that all magnitudes in the diluted regime exhibit square root of $ M_h $ {\bf scaling} laws and are 
{\bf universal} functions of $ r/r_h $ normalized to their values at the origin or at $ r_h $. 
Conversely, the halo mass $ M_h $ scales as the square of the halo radius $ r_h $ as
$$
 M_h = 1.75572 \; \Sigma_0 \; r_h^2 \quad .
$$
Moreover, the proportionality factor in this scaling relation is confirmed by the galaxy data (see fig. \ref{perfus}).

\medskip 

The phase space density decreases from its maximum value for the
compact dwarf galaxies corresponding to the limit of degenerate fermions till
its smallest value for large galaxies, spirals and ellipticals, corresponding to
the classical dilute regime. On the contrary, the halo radius $ r_h $ and the halo mass $ M_h $
monotonically increase from the quantum (small and compact galaxies) to the classical regime
(large and dilute galaxies).

Thus, the whole range of values of the chemical potential at the origin $ \nu_0
$ from the extreme quantum (degenerate) limit $ \nu_0 \gg 1 $ to the classical
(Boltzmann) dilute regime $ \nu_0 \ll -1 $ yield all masses, sizes, phase space
densities and velocities of galaxies from the ultra compact dwarfs till the
larger spirals and elliptical in agreement with the observations (see Table
\ref{pgal}).

\medskip

In addition, the galaxy velocity dispersions turn to be fully consistent 
with the galaxy observations in Table \ref{pgal} \cite{nos}.

\medskip

Adding baryons to CDM simulations have been often invoked to solve the
serious CDM problems at small scales. It must be noticed however that the
excess of substructures in CDM happens in DM dominated
halos where baryons are especially subdominat and hence the effects of
baryons cannot drastically modify the overabundance of substructures of the
pure CDM results.

The influence of baryon feedback into CDM cusps of density profiles
depends on the strength of the feedback. For normal values of the
feedback, baryons produce adiabatic contraction and the cusps in the density
profiles become even more cuspy.

Using the baryon feedback as a free parameter, it is possible
to exagerate the feedback such to destroy the CDM cusps
but then, the star formation ratio disagrees with the
available and precise astronomical observations.
Moreover, "semi-analytic (CDM + baryon)  models" have been introduced
which are just empirical fits and prescriptions to some galaxy observations.

In addition, there are serious evolution problems in CDM galaxies:
for instance pure-disk galaxies (bulgeless) are observed whose formation
through CDM is unexplained.

In summary, adding baryons to CDM simulations bring even more serious
discrepancies with the set of astronomical observations.

\medskip

We consider spherical symmetry in our approach for simplicity to determine 
the essential physical galaxy properties as the classical or 
quantum nature of galaxies, compact or dilute galaxies, 
the phase space density values, the cored nature of the mass density profiles, 
the galaxy masses and sizes \cite{nos,mnrpao,tgal}.  
It is clear that DM halos are not perfectly 
spherical but describing them as spherically symmetric is a first
approximation to which other effects can be added.
In ref. \cite{nos} we estimated the angular momentum 
effect and this yields small corrections. The quantum or classical 
galaxy nature, the cusped or cored nature of the density profiles in the 
central halo regions can be captured in the spherically symmetric treatment.

Our spherically symmetric treatment captures the essential features
of the gravitational dynamics and agree with the observations.
Notice that we are treating the DM particles quantum mechanically through
the Thomas-Fermi approach, so that expectation values are independent
of the angles (spherical symmetry) but the particles move and fluctuate
in all directions. Namely, this is more than treating purely classical orbits
for particles in which only radial motion is present. 
The Thomas-Fermi approach can be generalized to
describe non-spherically symmetric and non-isotropic situations,
by considering  distribution functions which include other 
particle parameters like the angular momentum.

\medskip

To conclude, the galaxy magnitudes: halo radius, galaxy masses and velocity dispersion
obtained from the Thomas-Fermi quantum treatment for WDM fermion masses in the keV scale are
fully consistent with all the observations for all types of galaxies (see Table \ref{pgal}). 
Namely, fermionic WDM treated quantum mechanically, as it must be, is able to reproduce
the observed DM cores and their sizes in galaxies \cite{nos,mnrpao,tgal}.
These results strenght the discussion in sec. \ref{qup} that compact galaxies are supported against 
gravity by the fermionic WDM quantum pressure. 

It is highly remarkably that in the context of fermionic WDM, the simple stationary
quantum description provided by the Thomas-Fermi approach is able to reproduce such broad variety of galaxies.

\medskip

Baryons have not yet included in the present study. This is fully justified for dwarf compact 
galaxies which are composed today 99.99\% of DM. In large galaxies the baryon fraction can
reach values up to  1 - 3 \%. Fermionic WDM by itself produces galaxies and structures in 
agreement with observations for all types of galaxies, masses and sizes. Therefore, the effect of including 
baryons is expected to be a correction to these pure WDM results, consistent with the fact that dark matter 
is in average six times more abundant than baryons.

\section{WDM gives the correct abundance of substructures}

It is known since some time through $N$-body simulations
that WDM alleviates the CDM satellite problem \cite{satel,satel2}
and the CDM voids problem \cite{tikho}.

{\vskip .2cm} 

WDM subhalos turns to be less concentrated than CDM subhalos.
WDM subhalos have the right concentration to host the bright
Milky Way satellites \cite{satel2}.

{\vskip .3cm} 

The ALFALFA survey has measured the velocity widths in galaxies from the 21cm HI line.
This is a test for substructure formation. The confrontation of the ALFALFA survey
with the substructures from $N$-body simulations clearly favours WDM over CDM \cite{simuwdm}.
A particle mass around $ \sim 2 $ keV is favoured by the ALFALFA survey.

\medskip

In summary, WDM produces the correct substructure abundance at zero redshift.

\medskip

Data on galaxy substructure for redshift $ z \lesssim 10 $ becomes now available.

{\vskip .2cm} 

In ref. \cite{mfl} the evolution of the observed AGN luminosity function for $ 3 < z < 6 $
is contrasted with the WDM and CDM simulations. 
WDM is clearly favoured over CDM by the observational data.

{\vskip .2cm} 

In ref. \cite{otroprd} the number of observed structures vs. the theoretical Press-Schechter estimation
for $ z =5, \; 6, \; 7 $ andf $ 8 $ is contrasted with the results from
WDM and CDM simulations. Again, WDM turns to be clearly favoured by the observations over CDM.

{\vskip .2cm} 

At intermediate scales where WDM and CDM give non-identical results and 
quantum effects are negligible, $N$-body classical simulations are reliable.
Contrasting such $N$-body classical simulations results with astronomical
observations at zero and non-zero redshifts {\bf clearly favours} WDM over CDM.

{\vskip .2cm} 

For larger scales $ \gtrsim 100 $ kpc, CDM and WDM $N$-body classical simulations are reliable
and give identical results in good agreement with astronomical and cosmological observations.

\section{Detection of keV mass Sterile Neutrinos}

Sterile neutrinos $ \nu_s $ are mainly formed by right-handed neutrinos $ \nu_R $ 
plus a small amount of left--handed neutrinos $ \nu_L $. Conversely, active neutrinos $ \nu_e $
are formed by $ \nu_L $ plus a small amount of $ \nu_R $:
$$ 
\nu_s \simeq \nu_R + \theta \; \nu_L \quad ,  \quad \nu_a = \nu_L + \theta \; \nu_R \; .
$$
The name sterile neutrino were was introduced by Bruno Pontecorvo in 1968.
They are singlets under all symmetries of the Standard Model of particle physics. 
Sterile neutrinos do not interact through weak, electro-magnetic or strong interactions.

\vskip 0.1cm

WDM $ \nu_s $ are typically produced in the early universe from active neutrinos through mixing,
namely, through a bilinear term $ \theta \; \nu_s \;  \nu_a $ in the Lagrangian.

\vskip 0.1cm

The appropriate value of the mixing angle $ \theta $
to produce enough sterile neutrinos $ \nu_s $ accounting for the observed total DM
depends on the particle physics model and is typically very small:
$$ 
\theta \sim  10^{-3}  -  10^{-4} \; .
$$ 
The smallness of $ \theta $ makes sterile neutrinos difficult to detect
in experiments.

{\vskip 0.2cm} 

Sterile neutrinos can be detected in beta decay and in electron capture (EC) processes
when a $ \nu_s $ with mass in the keV scale is produced {\bf instead} of an active $ \nu_a $:
$$
{}^3H_1 \Longrightarrow {}^3H \! e_{\, 2} + e^- + {\bar \nu}_e \quad ,
\quad {}^{187} Re \Longrightarrow {}^{187} Os  + e^- + {\bar \nu}_e  \; .
$$
In beta decays when a $ {\bar \nu}_s $ is produced instead of a 
$ {\bar \nu}_e $ in the decay products the electron spectrum is 
slightly modified at energies around the $ \nu_s $ mass ($\sim$ keV).
Such event can be inferred observing the electron energy spectrum. 
A 'kink' should then appear around the energy of the  $ \nu_s $ mass.

{\vskip 0.3cm} 

In electron capture processes like: 
$$
{}^{163}Ho  + e^- \Longrightarrow {}^{163}Dy^* + \nu_e \; ,
$$
when a sterile neutrino 
$ \nu_s $ with mass in the keV scale is produced {\bf instead} of an active $ \nu_e $,
the observed nonradiative de-excitation of the excited Dysprosium $ Dy^* $ is different
to the case where an active $ \nu_e $ shows up.

{\vskip 0.2cm} 
 
The available energies for these beta decays and EC are
\be\label{Q}
Q({}^{187}Re) = 2.47 \; {\rm keV} \quad , \quad  Q({}^3H_1) = 18.6 \; {\rm keV} \quad ,\quad  Q({}^{163}Ho) \simeq 2.5 \; {\rm keV}. 
\ee
In order to produce a sterile neutrino with mass $ m , \; Q $ must be larger than $ m $.
However, in order to distinguish the sterile neutrino $ \nu_s $ from a practically massless  
active neutrino $ \nu_a $,
$ Q $ must be as small as possible. This motivates the choice of the nuclei 
with the lowest known $ Q $ in eq.(\ref{Q}).

{\vskip 0.1cm}
 
For a theoretical analysis of $ \nu_s $ detection in Rhenium and Tritium beta decay 
see ref.\cite{renium} and references therein.

{\vskip 0.2cm}

Present experiments searching the small active neutrino mass also look for sterile
neutrinos in the keV scale:

\begin{itemize}
\item{MARE (Milan, Italy), Rhenium 187 beta decay and Holmiun 163 electron capture \cite{mare}.}
\item{KATRIN (Karlsruhe, Germany), Tritium beta decay \cite{katrin,kat2}.}
\item{ECHo (Heidelberg,  Germany), Holmiun 163 EC \cite{echo}.}
\item{Project 8 (Seattle, USA), Tritium beta decay \cite{p8}.}
\item{PTOLEMY experiment: Princeton Tritium Observatory.
Aims to detect the cosmic neutrino background and WDM (keV scale) sterile neutrinos through
the electron spectrum of the Tritium beta decay induced by the capture of a cosmic neutrino or
a WDM sterile neutrino \cite{pto}.}
\item{HOLMES electron capture in ${}^{163}$Ho calorimeter Gran Sasso}
\end{itemize}

The more popular sterile neutrino models nowadays are:

\begin{itemize}
\item{The Dodelson-Widrow (DW) model (1994): sterile neutrinos are produced by
non-resonant mixing from active neutrinos.}
\item{The Shi-Fuller model (SF) (1998): sterile neutrinos are produced by
resonant mixing from active neutrinos.}
\item{$\nu$MSM model (2005): sterile neutrinos are produced by
a Yukawa coupling from the decay of a heavy real scalar field $\chi$.}
\item{And in addition, there exists a variety of  sterile neutrino models based on the Froggatt-Nielsen mechanism, flavor symmetries,
$Q_6$, split see-saw, extended see-saw, inverse see-saw, loop mass. Furthermore: scotogenic,
LR symmetric models, etc. See for a recent review \cite{merle}.}
\end{itemize}

The primordial power spectra of WDM particles decoupling ultrarelativistically and out of equilibrium
in the first three WDM sterile neutrino models above (DW, SF and $\nu$MSM)
behave in a similar way just as if their
masses were different \cite{nosprd}. The masses of the WDM sterile neutrinos in the first three models which give the same
primordial power spectrum are related according to the formula \cite{nosprd}
(FD = thermal fermions):
 \be\label{conve}
\frac{m_{DW}}{\rm keV} \simeq 2.85 \; 
\left(\frac{m_{FD}}{\rm keV}\right)^{\! \frac43} \quad , \quad
m_{SF} \simeq 2.55 \;  m_{FD} \quad , \quad
m_{\nu{\rm MSM}} \simeq 1.9 \; m_{FD} \; .
\ee
(Here SF corresponds to the SF model without lepton asymmetry).

{\vskip 0.2cm}

The primordial spectra of DW, SF and $\nu$MSM models are equal between themselves
and equal to the themal relic power spectrum when the relations eq.(\ref{conve}) hold.

{\vskip 0.4cm}

On the other hand, sterile neutrinos $ \nu_s $ decay into active neutrinos $ \nu_a $ plus 
X-rays with an energy $ m/2 $ \cite{mp}. The lifetime of $ \nu_s $ is about
$ \sim 10^{11} \times $ age of the universe.
The value of the lifetime depends on the particle physics neutrino model.

{\vskip 0.1cm} 

These X-rays may be seen in the sky looking to galaxies \cite{kuse}.
See \cite{casey} for a recent review.

{\vskip 0.2cm} 

Some future observations of X-rays from galaxy halos are: 
\begin{itemize}
\item{DM bridge between M81 and M82 $ \sim 50$ kpc. Overlap of DM halos.
Satellite projects: Xenia (NASA) \cite{xenia}. ASTRO-H (Japan) \cite{astroh}.}
\end{itemize}

{\vskip 0.2cm} 

Some WDM hints from the CMB are:

\begin{itemize}
\item{WDM decay distorts the blackbody CMB spectrum.
The projected PIXIE satellite mission can measure the WDM sterile neutrino mass 
by measuring this distortion \cite{kogut}.}
\end{itemize}

{\vskip 0.1cm} 

Active neutrinos are very abundant in supernovae explosions and in these explosions
sterile neutrinos are produced too. Hence, bounds on the presence of
sterile neutrinos can be obtained contrasting to  supernovae observations.
The results from supernovae do not constrain $ \theta $ provided
$ 1 < m < 10 $ keV \cite{rz}.

\section{Sterile neutrinos and CMB fluctuations}

CMB fluctuations data provide the effective number of neutrinos, N$_{\rm eff}$.
This effective number  N$_{\rm eff}$ is related in a subtle way to the real number (three)
of active neutrinos plus the number of sterile neutrinos
with masses much smaller than the electron mass $ m_e $ \cite{advp}.

\medskip

WDM should decouple early at temperatures beyond the Fermi scale because
DM is not in the Standard Model of Particle Physics (SM). WDM couples 
 to the SM particles much weakly than weak interactions. Therefore, the number
of ultrarelativistic degrees of freedom at decoupling $ g(T_d) $ includes
all SM particles and probably beyond. We have $ g_{SM} = 427/4 $ for the SM and
$ g_{MSSM} = 915/4 $ for the minimal supersymmetric SM. 

{\vskip 0.1cm} 

Entropy conservation determines the WDM contribution to the effective number of neutrinos
N$_{\rm eff}$ \cite{kt}. One keV scale WDM sterile neutrino decoupling
at the temperature $ T_d $ contributes to N$_{\rm eff}$ at recombination by
\be\label{deltaN}
\Delta N^{WDM} = \left(\frac{T_d}{T_{rc}}\right)^4 = \left[\frac{g_{rc}}{g(T_d)}\right]^{4/3} 
\quad , \quad {\rm rc ~ stands ~ for ~ recombination} \; .
\ee
At recombination $ z=1090 $, we have $ g_{rc} = 29/4 $ and eq.(\ref{deltaN}) gives
for the SM and the MSSM:
$$
\Delta N^{WDM}_{SM} = 0.02771\ldots \quad , \quad \Delta N^{WDM}_{MSSM} = 0.01003\ldots
$$
Such keV WDM contributions to the effective number of neutrinos  N$_{\rm eff}$
are too small to be measurable by the present CMB anisotropy observations.
Hence, Planck and WMAP results cannot provide information about these keV
sterile neutrino WDM contributions.

{\vskip 0.1cm} 

However, Planck results \cite{pl} are {\bf compatible} with one or two Majorana sterile 
neutrinos in the eV mass scale \cite{advp}.
The possibility of a eV sterile neutrino is important for the whole subject 
of neutrinos and hence for WDM. The existence of one 
sterile neutrino in the eV scale opens the possibility of sterile neutrinos
suggesting the existence
of further sterile neutrinos with different masses, including a keV mass WDM sterile neutrino.

\section{Detection of a 3.56 keV X-ray line in galaxy clusters}

E.Bulbul et al.  \cite{bulbul} reported the detection of a new X-ray line 
in galaxy clusters that may be originated by the decay of a 7.1 keV sterile neutrino.
Sterile neutrinos remain out of thermal equilibrium today.

{\vskip 0.2cm} 

The line flux detected in the full sample corresponds to a
mixing angle for the decay  \cite{bulbul}
$$
\sin^2 2\; \theta \sim 7 \times 10^{-11}
$$
This mixing angle value is below the upper limits placed by the previous
searches.

{\vskip 0.2cm} 

From the conversion formulas eq.(\ref{conve}), a 7.1 keV DW sterile neutrino behaves as a 1.99 keV
thermal relic, and a 7.1 keV SF sterile neutrino behaves as a 2.8 keV thermal relic.
In addition, a 7.1 keV SF sterile neutrino with lepton asymmetry yields similar results \cite{aba}.

{\vskip 0.2cm} 

WDM thermal relics with thermal mass near 2 keV provide the correct small
scale structure formation and galaxy structures. Therefore, the main known sterile neutrino particle
models provide a 7.1 keV sterile neutrino and the same structure formation
results of a 2 keV thermal relic.

{\vskip 0.2cm} 

Therefore, a 7.1 keV sterile neutrino may be plausibily the dark matter particle !.

{\vskip 0.2cm} 

Confirmation of the detection and identification 
of the 3.56 keV X-ray line from Astro-H is  awaited for 2015 !

{\vskip 0.2cm} 

If a relic decoupling at thermal equilibrium would be the WDM, then
the favoured physical mass would be about 2 keV. This would require 
WDM particle models different from sterile neutrinos. 

\begin{figure}
\includegraphics[height=20.cm,width=20.cm]{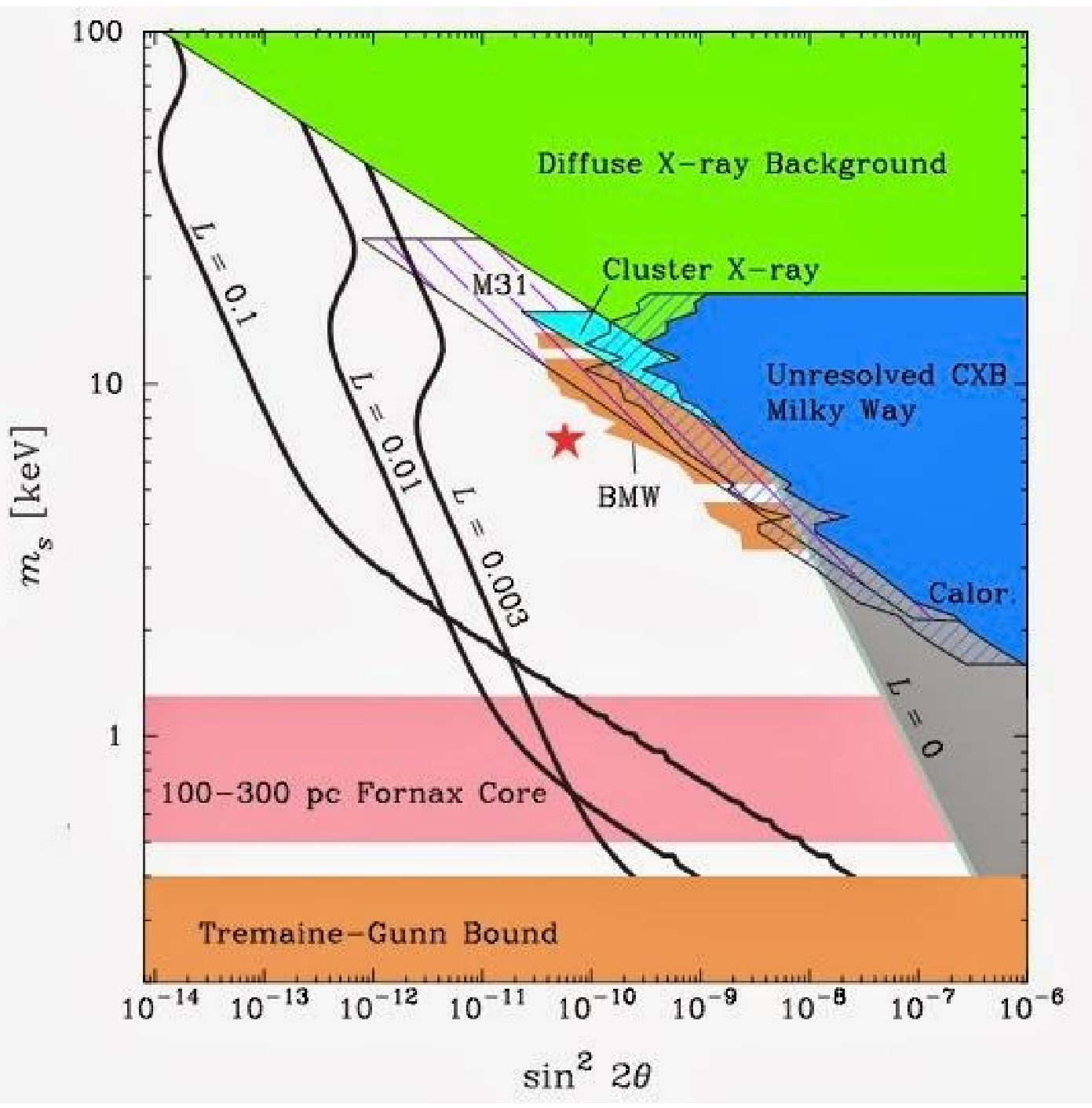}
\caption{Constraints on the sterile neutrino models from the literature. The
  full-sample line detection (assuming that the line is from sterile
  neutrino and that all dark matter is in sterile neutrino) is shown; error bar is statistical 90\%.  
  Historic constraints are taken from ref. \cite{aba09}. Black curves show
  theoretical predictions for the Dodelson-Widrow mechanism assuming that sterile
  neutrinos constitute the dark matter with lepton numbers L=0, 0.003, 0.01,
  0.1. See \cite{aba09} for explanation of the various observational
  constraints. 
The Bulbul et al. \cite{bulbul} detection of the sterile neutrino
indicated by a red star. The measurement lays at the boundary of the constraints from M31.}
\label{figbul}
\end{figure}

\section{Future Perspectives and Sterile Neutrino Detection}

WDM particle models must explain the baryon asymmetry of the universe.
This is a strong constraint on sterile neutrino models which must be
worked out for each model.

{\vskip 0.2cm} 

Combining particle, cosmological and galaxy results for sterile neutrinos 
at different mass scales \cite{nos,dvs,kopp,gni,drewes},
an appealing {\bf mass} neutrino hierarchy appears:
\begin{itemize}
\item{Active neutrino: $ \sim $ mili eV}
\item{Light sterile neutrino: $ \sim $ eV}
\item{Dark Matter sterile neutrino: $ \sim $ keV}
\item{Unstable  sterile neutrino: $ \sim $ MeV.... }
\end{itemize}
This scheme may represent the future extension of the Standard Model of particle physics.

{\vskip 0.1cm}

In order to falsify WDM, comprehensive theoretical calculations 
showing substructures, galaxy formation and evolution including 
the quantum WDM effects in the dynamical evolution are needed to contrast with
the astronomical observations.

{\vskip 0.1cm}

In such WDM theoretical calculations the quantum pressure must be necesarily included.
These calculations should be performed matching the semiclassical
Hartree-Fock (Thomas-Fermi) dynamics where the dimensionless phase-space
density is high enough, namely,  $ \hbar^3 \; Q/m^4 \gtrsim 0.1 $ with the
classical evolution dynamics where $  \hbar^3 \; Q/m^4 \ll 1 $. 
These are certainly not easy numerical calculations but they are unavoidable!

{\vskip 0.2cm}

Richard P. Feynman foresaw the necessity to include quantum physics in simulations in 1981 \cite{fey}

\begin{center}

{\vskip 0.2cm}

{\bf ``I'm not happy with all the analyses that go with just the classical theory, because nature isn't classical, dammit, and if you want to make a simulation of nature, you'd better make it quantum mechanical, and by golly it's a wonderful problem, because it doesn't look so easy.'' }
\end{center}

Sterile neutrino detection depends upon the particle 
physics model. There are sterile neutrino models where the
keV sterile is stable and thus hard to detect (see for example \cite{merle}).

{\vskip 0.2cm}

Detection may proceed through 
astronomical observations of X-ray keV sterile neutrino decay
from galaxy halos and by direct detection of sterile neutrinos in 
laboratory experiments.

{\vskip 0.2cm}

Mare \cite{mare}, Katrin \cite{katrin,kat2}, ECHo \cite{echo},  Project 8 \cite{p8}, HOLMES and Ptolemy \cite{pto}
are expected to provide
bounds on the mixing angles. However, for a direct particle detection, a 
dedicated beta decay experiment and/or electron capture experiment seem necessary
to find sterile neutrinos with mass in the keV range.
In this respect, calorimetric techniques seem well suited.

{\vskip 0.1cm}

The best nuclei for study are ${}^{187}$Re and Tritium for beta decay and ${}^{163}$Ho for electron capture.
However, only Tritium has enough available energy $ Q $ to produce 7 keV sterile neutrinos.

{\vskip 0.3cm}

The search of DM particles with mass around 7 keV is a promisory
avenue for future trascendental discoveries.

\end{document}